\documentclass[twocolumn,secnumarabic,amssymb, nobibnotes, aps, prd]{revtex4}
\usepackage{graphicx}

\begin{document}

\narrowtext

{\bf Comment on ``Probing Noise in Flux Qubits via Macroscopic Resonant Tunneling''}

 \bigskip

The authors of the paper \cite{Harris} consider superconducting loop interrupted by Josephson junctions as flux qubit and interpret their experimental results as macroscopic resonant quantum tunneling between macroscopically distinct quantum states. But this interpretation can not be correct because of an irreconcilable contradiction with the law of angular momentum conservation. The persistent current has opposite direction (counter-clockwise in $|0>$ and clockwise in $|1> $, see Fig.1 in \cite{Harris}) and consequently different magnetic moment $M_{m} = I_{p}S$ and angular momentum of Cooper pairs $M_{p} = (2m_{e}/e)M_{m}$ in the two states between which the quantum tunnelling is assumed. Thus, the authors \cite{Harris} assume a causeless macroscopic change of magnetic moment $\Delta M_{m} = I_{p}S - (-I_{p}S) = 2I_{p}S$ and angular momentum $\Delta M_{p} = (2m_{e}/e)\Delta M_{m}$. $\Delta M_{m} \approx  10^{5} \ \mu _{B}$ and $\Delta M_{p} \approx 10^{5} \ \hbar $ at the persistent current $|I_{p}| \approx  0.5 \ \mu A $ and loop area $S \approx  1 \ \mu m^{2} $ of a typical ``flux qubit''. According to the universally recognized quantum formalism such causeless change is not possible even for microscopic magnitude $\Delta M_{p} = \hbar  $ because of its contradiction with the conservation law. 

I should note that the application by many authors of the spin- 1/2 formalism developed for a central-symmetrical object \cite{Landau} to ``flux qubit'' states misleads. The loop, in contrast to electron, is no central-symmetrical object. The superposition of the z- component eigenstates of the spin-1/2 
$$\Psi = \cos(\frac{\theta }{2}) |0>_{z} + \sin(\frac{\theta }{2})  \exp \frac{\phi }{2}|1>_{z} \eqno{(1)} $$
presupposes eigenstate of a component along a direction rotated relatively $z$ on angles $\theta $ and $\phi $. We can change the co-ordinate system for central-symmetrical object and reduce the superposition (1) to an eigenstate. For example at $\theta = \pi/2$, i.e. $cos(\theta /2) = sin(\theta /2) = \surd {2}$, we can designate the direction corresponding to the eigenstate as $y$  ($\phi = \pi/2$) and reduce the superposition (1) to the z-eigenstate $\Psi  = |0>_{z}$ using the rotation operator \cite{Landau} 
$$U_{x}(\theta ) = \cos(\frac{\theta }{2}) +i\sigma _{x}\sin(\frac{\theta }{2})  \eqno{(2)} $$
with $\theta =-\pi /2$, where $\sigma _{x}$ is the Pauli operator. According to the universally recognized quantum formalism the eigenstate $\Psi  = |0>_{z}$ means that a measurement of this spin component should give with certainty, i.e. with probability equal to unity, positive value $+1/2\hbar $. 

But we can not change the co-ordinate system for the ``flux qubit''. Therefore the effective Hamiltonian (1) in  \cite{Harris} with the Pauli matrices misleads. It must presuppose that there is a probability to measure non-zero component of $M_{p}$ or $M_{m}$ in the plane of the ``flux qubit'' loop. Any physicists must understand that this can not be correct since a circular current $I_{p}$ flowing in a plane induces no magnetic moment component in this plane. The magnetic moment of the ``flux qubit'' $|M_{m}| \approx 0.5 \ 10^{5} \  \mu _{B}$ is one-dimensional in contrast to the one $M_{m} =  \mu _{B}$ of electron.  

The universally recognized quantum formalism excludes the violation of the law of the angular momentum conservation thanks to the von Neumann's projection postulate (collapse of the wave function at measurement) \cite{Neumann}. The superposition (1) must collapse to an eigenstate, for example $\Psi  = |0>_{z}$, at first measurement of the z -component and all posterior measurements of the z- component must give with certainty the same result $+1/2\hbar  $ when any interaction with environment is absent. Therefore the interpretation of the change of angular momentum of Cooper pairs from $M_{p} \approx 0.5 \ 10^{5} \ \hbar $ to $M_{p} \approx -0.5 \ 10^{5} \ \hbar $ with a small variation of the externally applied magnetic flux from $BS =\Phi = (n + 1/2 - \delta )\Phi _{0}$ to $\Phi = (n + 1/2 + \delta )\Phi _{0}$ (the typical interval $\delta  \approx  0.005$ and $2\delta B =2\delta \Phi _{0}/S  \approx  2 \ 10^{ - 5} \ T$ at $S \approx 10^{ - 12} \ m^{2}$) observed at measurements of the ``flux qubit'' persistent current \cite{Mooij} as a consequence of quantum tunneling or superposition of two macroscopically distinct quantum states contradicts not only to the law of the angular momentum conservation but also to the universally recognized quantum formalism. 

The mass delusion concerning superposition of superconducting loop states manifested in \cite{Harris} may be a consequence of misunderstanding of the matter of quantum description by many modern physicists. In order to overcome the prohibition of the quantum tunnelling between states with different angular momentum some authors assume a firm coupling with a large solid matrix that absorbs the change in the angular momentum \cite{Chudnovsky}. This assumption presupposes that quantum tunnelling and superposition of states are possible not only for Cooper pairs but also the loop, substrate and so forth. Such quantum fantasy has nothing in common with the universally recognized quantum formalism. Its creators understood that quantum theory would provide predictions concerning the results of measurements, but, unlike all previous theories, it is incapable of providing a full account of {\it "how nature did it"}. Niels Bohr stated: ``{\it There is no quantum world. There is only an abstract quantum physical description}'', the citation from \cite{Zeilinger}. According to this positivism point of view there is no sense to dream up about {\it a firm coupling with a large solid matrix that absorbs the change in the angular momentum} \cite{Chudnovsky}. Superposition and quantum tunnelling describe only phenomena, i.e. results of observations, but no a real physical process. Therefore the assumption \cite{Harris} on quantum tunnelling and superposition of the ``flux qubit'' states presupposes unambiguous violation of the law of the angular momentum conservation at observation, i.e. in phenomena, which are the matter of {\it the abstract quantum physical description}.

The famous paradoxes proposed as far back as 1935 \cite{EPR,Schrodinger} have accentuated that phenomena described with help of state superposition and quantum tunnelling contradict to local realism \cite{EPR} and that they are causeless \cite{Schrodinger}. The assumption on superposition of macroscopically distinct quantum states contradicts also to macroscopic realism \cite{Legget85}. It is strange that many authors venture to claim about realism failure on the macroscopic level basing only on experimental results obtained at measurements of two-state quantum system in defiance of the Bell's hidden-variables model for a single spin -1/2 that reproduces all predictions of measurement results given by the orthodox quantum theory \cite{Bell66,Mermin93}. 

The ``Schrodinger cat'' paradox \cite{Schrodinger} is associated by many authors \cite{Cat,QCh} with the problem of macroscopic quantum phenomena only because that all cats, which we know, are macroscopic. But it is obvious that nothing in this paradox could depend on size of the cat. Nothing, except tragedy situation, could change in this paradox at substitution of cat, small flask of hydrocyanic acid and hammer for a recorder which can record the discharge of Geiger counter tube. The ``Schrodinger cat'' emphasizes paradoxicality of no macroscopic quantum phenomena but of quantum indeterminacy of phenomena describing with quantum tunnelling and state superposition. Schrodinger in his paradox \cite{Schrodinger} presupposed a cause - effect connection between the states of the small flask and the cat, the Geiger counter tube and the flask, the atom and the Geiger counter tube (see the short Schrodinger's description on page 185 of \cite{QCh}). When we will open the steel chamber with the cat, the $\Psi $-function describing superposition of states should collapse. If we see, for example, that the cat is dead we can conclude that the cat is dead since the hammer has shattered the flask with poison. The hammer has shattered the flask since the Geiger counter tube has discharged. The Geiger counter tube has discharged since the atom has decayed. Till now each result had the cause. But nobody can say why the atom has decayed. Just this absent of the cause is described by superposition of states and quantum tunnelling. The quantum formalism assumes that phenomena can be causeless. But a causeless change of angular momentum or other violation of conservation laws are considered to be inadmissible.

Because of the contradiction with the local realism many experts force to think \cite{Heisenberg} that ``{\it the wavefunction or, more generally, the density matrix represent our knowledge of the system we are trying to describe}'' (the citation from \cite{Peierls}). This information interpretation saves from reality of quantum paradoxes connected with the collapse of the Schrodinger wave function at observation. But it can not be universal since the wave function describing the real density of Cooper pairs can not represent our knowledge \cite{FPP5} in contrast to the Schrodinger wave function describing a probability. It is important to note that this wavefunction can not collapse since the real density of Cooper pairs can not change because of our look \cite{FPP5}. This difference in essence means that the wavefunction describing quantum phenomena in superconductor can not be used for description of state superposition. Therefore an new additional wavefunction, which can collapse, is fabricated in \cite{Harris} and many other publications for the description of the ``flux qubit'' state superposition. This new wavefunction can describe nothing since phenomena described with it must violate the law of angular momentum conservation. 

 \bigskip
A.V. Nikulov

Institute of Microelectronics Technology 

and High Purity Materials, 

Russian Academy of Sciences, 

142432 Chernogolovka, Moscow District, 

RUSSIA.

\end{document}